\documentclass[proof]{pasj00}
\usepackage{color}

\begin{document}
\SetRunningHead{S. Notsu et al.}{High Dispersion Spectroscopy of the Superflare Star KIC6934317}
\Received{2013 February 8}
\Accepted{2013/07/18}

\title{High Dispersion Spectroscopy of the Superflare Star KIC6934317 \thanks{Based on data collected at Subaru Telescope, which 
is operated by the National Astronomical Observatory of Japan.}}

\author{Shota \textsc{Notsu}\altaffilmark{1}, Satoshi \textsc{Honda}\altaffilmark{2,3}, Yuta \textsc{Notsu}\altaffilmark{1}, 
Takashi \textsc{Nagao}\altaffilmark{1}, Takuya \textsc{Shibayama}\altaffilmark{1}, Hiroyuki \textsc{Maehara}\altaffilmark{2,4},  
Daisaku \textsc{Nogami}\altaffilmark{2}, and Kazunari \textsc{Shibata}\altaffilmark{2}}

\affil{\altaffilmark{1}Department of Astronomy, Faculty of Science, Kyoto University, Kitashirakawa-Oiwake-cho,
Sakyo-ku, Kyoto 606-8502}
\affil{\altaffilmark{2}Kwasan and Hida Observatories, Kyoto University, Yamashina-ku, Kyoto 607-8471}
\affil{\altaffilmark{3}Center for Astronomy, University of Hyogo, 407-2, Nishigaichi, Sayo-cho, Sayo, Hyogo 
679-5313}
\affil{\altaffilmark{4}Kiso Observatory, Institute of Astronomy, School of Science, The University of Tokyo, 10762-30, Mitake, 
Kiso-machi, Kiso-gun, Nagano 397-0101}
\email{snotsu@kwasan.kyoto-u.ac.jp }


\maketitle

\begin{abstract}
\noindent We conducted the high-resolution spectroscopic observation with Subaru/HDS for a G-type star (KIC6934317). We
selected this star from the data of the Kepler spacecraft. This star produces a lot of superflares, and
the total energy of the largest superflare on this star is $\sim 10^{3}$ times larger ($\sim 2.2\times 10^{35}$ erg)
than that of the most energetic flare on the Sun ($\sim 10^{32}$ erg).
The core depth and emission flux of Ca II infrared triplet lines and H$\alpha$ line 
show high chromospheric activity in this star, in spite of its low lithium abundance 
and the small amplitude of the rotational modulation.
Using the empirical relations between 
emission flux of chromospheric lines and X-ray flux, this star is considered to show much higher coronal activity than 
that of the Sun.
It probably has large starspots which can store a large amount of magnetic energy enough to give rise to superflares. 
We also estimated the stellar parameters, such as effective temperature, surface gravity, metallicity, projected rotational 
velocity ($v \sin i$), and radial velocity.
KIC6934317 is then confirmed to be an early G-type main sequence star.
The value of $v \sin i$ 
is estimated to be $\sim 1.91$ km $\mathrm{s}^{-1}$.
In contrast, the rotational velocity is calculated to be $\sim$20 km $\mathrm{s}^{-1}$ by using the period of the brightness variation as the 
rotation period.
This difference can be explained by its small inclination angle (nearly pole-on).
The small inclination angle is also supported by the contrast between the large 
superflare amplitude and the small stellar brightness variation amplitude.
The lithium abundance and isochrones implies that the age of this star is more than about a few Gyr, though 
a problem why this star with such an age has a strong activity remains unsolved.

\end{abstract}

\section{Introduction}
\noindent Flares are explosions on the stellar surface, and are thought to occur by intense release of 
magnetic energy stored near starspots (e.g., \cite{Shibata2011}). 
The typical amount of energy released in solar flares is $10^{29}-3\times 10^{32}$ erg (e.g., \cite{Priest1981}; 
\cite{Shibata2002}).
\\ \\
\citet{Schaefer2000} found 9 candidates of superflares on slowly rotating stars like the Sun.  Superflares are the flares 
whose total energy is $10-10^6$ times larger ($\sim 10^{33}-10^{38}$ erg) than that of the most energetic flare on the Sun 
($\sim10^{32}$ erg). Recently, \citet{Maehara2012}
discovered 365 superflare events on 148 solar-type stars that have effective temperature of 
5100K $\leq$ $T_{\mathrm{eff}} <$ 6000K, and surface gravity of $\log g$ $\geq$ 4.0 by analyzing the data of the Kepler 
spacecraft (\cite{Koch2010}). In addition, \citet{Shibayama2013} discovered 1547 superflares on 279 solar-type stars.
The Kepler spacecraft is particularly suitable for the detection of flares of small intensity, against the integrated 
star disk brightness, because of the combination of high photometric accuracy ($10^{-6}$ mag) and continuous time-series data of 
a lot of stars over a long period (e.g., \cite{Walkowicz2011}; \cite{Balona2012}).
\\ \\
\citet{Maehara2012} and \citet{Shibayama2013} suggested that superflares releasing a total energy in the range 
$\sim 10^{34}-10^{35}$ erg can occur once in $800-5000$ years on Sun-like stars that have effective temperature of 
5600K $\leq$ $T_{\mathrm{eff}} <$ 6000K, surface gravity of $\log g$ $\geq$ 4.0, and the rotational period of $P$ $\geq$ 10 days.
Many of solar-type stars having superflares show quasi-periodic brightness variations with a typical period from one day to a 
few tens of days. The amplitude of the brightness variations is in the range 0.1$\sim$10$\%$ 
(\cite{Maehara2012}; \cite{Shibayama2013}).
Such brightness variations can be explained by the rotation of the star having starspots (e.g., \cite{Basri2011}; 
\cite{Debosscher2011}; \cite{Harrison2012}). 
\\ \\
\citet{YNotsu2013} found that the brightness variations of superflare-generating stars can be well 
explained by the rotation of the star with fairly large starspots, taking into account the effects of inclination angle and spot 
latitude. They also confirmed the correlation between the starspot coverage and the maximum energy of superflares.
These results indicate that the energy of superflares can be explained by the magnetic energy stored around the starspots.
In addition, \citet{Shibata2013} suggested from theoretical estimates that the Sun can generate 
large magnetic flux which is enough for causing superflares with energy of $10^{34}$ erg 
within one solar cycle period ($\sim$ 11 years).
\\ \\
If such large starspots exist on a star, the activity level of its chromosphere becomes high and large plages occur near 
the starspots (\cite{Shine1972}).
\citet{Linsky1979} and \citet{Foing1989} suggested that the lines of Ca II infrared  triplet (Ca II IRT, $\lambda$ = 8498, 8542, 8662\AA) 
indicate the chromospheric activity of G-type stars. This is because these are collision-dominated lines and their
cores are formed in the chromosphere, reflecting the temperature rise in their profiles which go from filled-in to pure emission, 
depending on the activity level.
The fluxes in these lines are known to correlate well with the Ca II H+K emission index (log$R_{\mathrm{HK}}'$) and emission flux, which has been used as the 
indicator of chromospheric activity more traditionally (e.g., \cite{Rutten1984}; \cite{Schrijver1992}; 
\cite{Chmielewski2000}; \cite{Busa2007}; \cite{Hall2008}; \cite{Martinez-Arnaiz2011}; \cite{Takeda2012}).
Solar imaging spectroscopy in these lines is also conducted to investigate chromospheric structures, such as plages and fibrils 
(e.g., \cite{Shine1972}; \cite{Cauzzi2008}; \cite{Reardon2009}).
\\ \\
In addition, \citet{Herbig1985}, \citet{Soderblom1993}, and \citet{Frasca2010} suggested that the H$\alpha$ line also indicates the chromospheric 
activity of the Sun and G-type stars. This is because the optical depth in the line core is large, 
and because the line formation becomes more and more dominated by collisions in very active stars or in plage regions.
Therefore, it is very important to conduct high-resolution spectroscopic observations to investigate the chromospheric activity
 of stars generating superflares, which is related to their photospheric starspots.
\\ \\ 
In this research, we analyze the high-resolution spectrum of a G-type superflare star (KIC6934317), 
and investigate the chromospheric activity and the presence of starspots.
We also estimate stellar parameters of the star, such as $T_{\mathrm{eff}}$, $\log g$,  
metallicity represented by the Fe abundance relative to 
the Sun ([Fe/H])\footnote[1]{ [Fe/H] is presented by the following relation,
 
 \begin{equation}
 \textnormal{[Fe/H]} = \log (\mathrm{N}_{\mathrm{Fe}}/\mathrm{N}_{\mathrm{H}})_{\mathrm{star}} - \log (\mathrm{N}_{\mathrm{Fe}}/\mathrm{N}_{\mathrm{H}})_{\mathrm{sun}},
 \end{equation}
 
\noindent where $(\mathrm{N}_{\mathrm{Fe}}/\mathrm{N}_{\mathrm{H}})_{\mathrm{star}}$ is the ratio of the 
number of Iron (Fe) atoms to that of Hydrogen (H) atoms 
in the star, and $(\mathrm{N}_{\mathrm{Fe}}/\mathrm{N}_{\mathrm{H}})_{\mathrm{sun}}$ is the same ratio 
in the Sun. \citet{Asplund2009} reported that $\log (\mathrm{N}_{\mathrm{Fe}}/\mathrm{N}_{\mathrm{H}})_{\mathrm{sun}}$ is 
$-4.5\pm 0.04$. }, projected rotational 
velocity ($v\sin i$) and the radial velocity (RV), and 
discuss some features of this star.
This is the first high-resolution spectroscopic study of G-type superflare-generating stars discovered by Kepler spacecraft.
\\ \\ 
We discuss the selection of the target and our high-resolutions spectroscopic 
observation in Section \ref{sec:Targets and Observation}. 
We show the method to measure stellar parameters in Section \ref{sec:Measuments pf Stellar Parameters}.
The results of stellar parameters, such as chromospheric parameters and $v \sin i$, are reported in Section \ref{sec:Results}.
Finally in Section \ref{sec:Discussion}, we discuss the stellar activity, the rotational velocity, and the binarity of this star.
\\ \\

\section{Targets and Observation}
\label{sec:Targets and Observation}

\subsection{Target Selection}
\noindent We selected the G-type superflare star (KIC6934317) as the target. This star exhibited 48 
superflares in about 617 days, and hence the superflare occurrence frequency of this star is about once 
in 13 days. We selected the star from the data 
of the Kepler spacecraft (\cite{Koch2010}) within the temperature range between 5100K to 6000K.
These Kepler data were retrieved 
from the Multimission Archive at Space Telescope Science Institute (MAST)\footnote[2]{http://archive.stsci.edu/kepler/}. 
In addition, we applied the analysis methods described in \citet{Maehara2012} and \citet{Shibayama2013} to search for superflares. 
According to KIC (Kepler Input Catalog, \cite{Brown2011}), this star is 12.03 mag in the $i$ band and 
was the brightest superflare star at present in our data. 
Table 1 shows the photometric data of KIC6934317 taken from previous studies in detail. 
The $\textit{UBV}$ magnitudes are taken from the photometric survey of the 
Kepler Field\footnote[3]{The data of this survey is 
available at ~ $\mathrm{http://archive.stsci.edu/kepler/kepler \_fov/search.php/}$ .} (\cite{Everett2012}),  
and those of $JHK_{\mathrm{s}}$ magnitudes 
are from the data of Two Micron All Sky Servey (2MASS; \cite{Skrutskie2006}).
The atmospheric parameters reported in the KIC (\cite{Brown2011}) are $T_{\mathrm{eff}} = 5387\pm 200$K, $\log g = 3.8\pm 0.4$.
They also quote a radius of 2.3$\RO$.
\\ \\
Figure 1(a) shows the light curve of KIC6934317 and reveals a quasi-periodic brightness variation. The amplitude of 
this variation is small ($\sim 0.1 \% $) compared to those of other stars which undergo superflares. 
Figure 1(b) is an enlarged lightcurve of a superflare observed around BJD 2455257.9. 
The amplitude of this flare is about 2.1\% of the brightness of this star, the duration of the flare 
is about 0.12 day, and the total radiative energy during this event is
about $5.6 \times 10^{34}$ erg.
\citet{Maehara2012} and \citet{Shibayama2013} described the method of estimating the total energy 
of each flare in detail.
The largest superflare on KIC6934317 was observed around BJD 2455735.3.
Its amplitude is about 5.7\% of the brightness of this star, the duration of this flare is about 0.10 day, 
and the total radiative energy released during this event is about $2.2 \times 10^{35}$ erg.
Figure 1(c) shows the power spectrum of the time variation of the stellar brightness of KIC6934317 
and it shows that the period of the brightness variation ($P_{\mathrm{s}}$) is about 2.54 days.
\\ \\
We also selected 59Vir and 61Vir as references of G-type stars.
Previous studies (e.g., \cite{Anderson2010}) have shown that 59Vir rotates rather fast and has strong 
magnetic fields ($\sim 500$G), while 61Vir is slowly rotating and no magnetic field could be detected. 
\\ \\
Table 2 shows in detail some stellar parameters (atmospheric parameters, projected rotational velocity, radial velocity, and lithium abundance) of KIC6934317 taken from previous studies and derived 
in this study, and Table 3 shows those of our comparison stars (59Vir, 61Vir).
\\ \\

\subsection{Subaru Observations and Data Reduction}
\noindent We carried out high dispersion spectroscopy of KIC6934317, 59Vir, and 61Vir on August 3, 2011 
(Hawaii Standard Time) in Subaru service program in semester 11B (S11B-137S). 
We used High Dispersion Spectrograph (HDS: \citet{Noguchi2002}) at the 8.2-m Subaru telescope. 
The spectral coverage was about 6100\AA$\sim$8820\AA. This range includes important lines, such as those 
of Ca II infrared triplet 
(8498, 8542, 8662\AA), H$\alpha$ (6562.8\AA), the Li I line (6708\AA), and some lines of Fe I
and Fe II (for measuring atmospheric parameters, projected rotational velocity, and radial 
velocity). The exposure time of KIC6934317 was 
$1800\times 6$ seconds to get signal-to-noise (S/N) ratio as $\sim$150 at 8520\AA \ (around the Ca II IRT lines), 
and as $\sim$210 at 6700\AA \ (around the line of Li I 6708\AA). 
The exposure times of 59Vir and 61Vir were 
$60\times 2$ seconds per each star. S/N ratio of 59Vir is $\sim$ 370 at 8520\AA \ and $\sim$ 540 at 6700\AA \, and that of 61Vir is $\sim$
410 at 8520\AA \ and $\sim$690 at 6700\AA.
\\ \\
The spectral resolution, as evaluated from the full width at half maximum (FWHM) of the emission lines of the Th-Ar calibration lamp, was 
0.088\AA \ at 8500\AA, yielding a resolving power R=$\lambda$/$\Delta$$\lambda$$\sim$97,000 (for the slit width of 0.36 arcsec).
\\ \\
The data reduction (bias subtraction, flat fielding, aperture determination, scattered light subtraction, spectral extraction, wavelength calibration,
and normalization by the continuum) was performed with the ECHELLE package of the IRAF\footnote[4]{IRAF is distributed by the National Optical 
Astronomy Observatories, which is operated by the Association of Universities for Research in Astronomy, Inc., under cooperate agreement with 
the National Science Foundation.} software.
\\ \\
Figure 2 shows the lightcurve of KIC6934317 around the period of this observation (around BJD 2455778). 
The spectrum was taken during the minimum of the wave-like modulation.
There was a large flare about a day before the observation period.
\\ \\

\section{Measurements of Stellar Parameters}
\label{sec:Measuments pf Stellar Parameters}

 \subsection{Atmospheric Parameters}
 \noindent We measured equivalent widths of 101 Fe I lines and 7 Fe II lines over the range of 
 $6100-7400$\AA \ of the target stars (KIC6934317, 59Vir, and 61Vir).
 These Fe I and II lines are selected from the line list presented in \citet{Takeda2005b}, which is based 
 on many previous studies (e.g., \cite{Kuruxz1984}; \cite{Meylan1993}; \cite{Kurucz1995}). 
 In this process of measuring equivalent widths, we used the code SPSHOW contained in 
 SPTOOL software package\footnote[5]{http://optik2.mtk.nao.ac.jp/$\sim$takeda/sptool/} developed 
 by Y. Takeda. Programs contained in SPTOOL are 
 originally based on Kurucz's ATLAS9/WIDTH9 model atmospheric programs (\cite{Kurucz1993}).
 \\ \\
 Using these data of equivalent widths of Fe I and Fe II lines, we derived the effective temperature ($T_{\mathrm{eff}}$), 
 the surface gravity ($\log g$), microturbulent velocity ($v_{\mathrm{t}}$), and [Fe/H] of target stars.
 For deriving these parameters, we used TGVIT 
 program\footnote[6]{http://optik2.mtk.nao.ac.jp/$\sim$takeda/tgv/} developed by Y. Takeda.
 The procedures adopted in this program are minutely described in \citet{Takeda2002} and \citet{Takeda2005b}.
 \\ \\
 The resultant parameters of $T_{\mathrm{eff}}$, $\log g$, $v_{\mathrm{t}}$, and [Fe/H] of 
 KIC6934317 are $5694\pm 25$K, 
 $4.42\pm 0.08$, $0.87\pm 0.14$ km $\mathrm{s}^{-1}$,  and $-0.03\pm 0.07$, respectively. 
 On the other hand, the atmospheric parameters reported in the KIC (\cite{Brown2011}) are $T_{\mathrm{eff}} = 5387\pm 200$K, 
 $\log g = 3.8\pm 0.4$, and [Fe/H] $=-0.78\pm 0.5$. There are significant differences between the values 
 reported in the KIC and those derived in this observation. 
 However, the atmospheric parameters reported in the KIC are derived from 
 multiband photometry for a first classification and are not good sources to discuss actual properties of the stars.
 Furthermore, it is known that in some cases the atmospheric parameters taken from the KIC are significantly
 different from those derived spectroscopically, which are much more accurate than those from the 
 KIC (e.g., \cite{Molenda-Zakowicz2010}, \cite{Pinsonneault2012}).
 \\ \\
 In addition, the resultant parameters of 59Vir are $T_{\mathrm{eff}}=6009\pm 28$K, $\log g=4.15\pm 0.06$, 
 $v_{\mathrm{t}}=1.32\pm 0.09$ km $\mathrm{s}^{-1}$, and [Fe/H]$=0.09\pm 0.06$, and 
 those of 61Vir are $T_{\mathrm{eff}}=5558\pm 15$K, $\log g=4.50\pm 0.04$, 
 $v_{\mathrm{t}}=0.87\pm 0.08$ km $\mathrm{s}^{-1}$, and [Fe/H]$=-0.04\pm 0.06$. These values are consistent with those of previous 
 studies (e.g., \cite{Takeda2007}; \cite{Schroder2009}; \cite{Anderson2010}).
 Obviously, the errors of the parameters are internal to the procedure and do not 
 include any systematic effect due to the choice of the line-list and the model atmosphere.
 \\ \\
 For the sake of checking the accuracy of the atmospheric parameters derived from equivalent widths of Fe I and Fe II lines, we also estimated  
 $T_{\mathrm{eff}}$ from the ratio of the depth of two different lines (e.g., \cite{Biazzo2007}).
 We measured the line-depth ratio of three pairs of lines (V I 6199\AA/Fe I 6200\AA, V I 6214\AA/Fe I 6213\AA, 
 and V I 6275\AA/Fe I 6270\AA), and roughly estimated the average values of $T_{\mathrm{eff}}$ by using the method 
 described in \citet{Biazzo2007}.
 The value of $T_{\mathrm{eff}}$ of KIC6934317, 59Vir and 61Vir is $\sim$5850K, $\sim$6060K, and $\sim$5800K, respectively. 
 These values are roughly consistent with the values of $T_{\mathrm{eff}}$ derived from equivalent widths of Fe I and Fe II lines.
 \\ \\
 \citet{Pinsonneault2012} reported a catalog\footnote[7]{This catalog is available 
 at \ http://vizier.cfa.harvard.edu/viz-bin/VizieR-3 .} of revised $T_{\mathrm{eff}}$ for stars in the KIC (\cite{Brown2011}).
 They estimated $T_{\mathrm{eff}}$ by using two different methods and compared their values with those in the KIC.
 In one method, they transformed original $griz$ colors in the KIC into $griz$ colors on the basis of 
 Sloan Digital Sky Survey 
 (SDSS; \cite{Aihara2011}) scales and calculated $T_{\mathrm{eff}}$ from the relations between SDSS $griz$ colors and 
 $T_{\mathrm{eff}}$. Hereafter, we refer to this method as the SDSS one.
 In the other method, $T_{\mathrm{eff}}$ is derived with the infrared flux method (IRFM; e.g., \cite{Casagrande2010}) which uses
 the relations between $T_{\mathrm{eff}}$ and $J-K\mathrm{s}$ 
 from the data of 2MASS (\cite{Skrutskie2006}). We call this the IRFM method. The effects of interstellar reddening are also corrected 
 in these two methods of estimating $T_{\mathrm{eff}}$ (\cite{Casagrande2010}, \cite{Pinsonneault2012}).
 They argued that these revised values of $T_{\mathrm{eff}}$ with two different methods are both about 200K higher than the values
 of $T_{\mathrm{eff}}$ in the KIC,  and these revised values are comparable with the values derived spectroscopically.
 \\ \\
 According to this catalog, the revised $T_{\mathrm{eff}}$ of KIC6934317 by using the SDSS method is $5710\pm 81$K, and the 
 value with the IRFM method is $5604\pm 103$K. These revised values are higher than that in the KIC, and are consistent 
 with the result of this spectroscopic observation within the error range.
 \\ \\
 On the basis of results derived spectroscopically, KIC6934317 is an early G-type main sequence star, and 
 $T_{\mathrm{eff}}$, $\log g$, and [Fe/H] of this star are nearly the same as the Sun.
 Atmospheric parameters of KIC6934317 taken from previous studies and derived in this study are shown in Table 2, and
 those of comparison stars (59Vir, 61Vir) are shown in Table 3.
 \\ \\

 \subsection{$v \sin i$}
 \noindent We used the method which is basically similar to that described in \citet{Takeda2008} in order to
 derive $v \sin i$ of KIC6934317, the Sun, 59Vir, and 61Vir.
 In this process of calculating $v \sin i$, we took into account the 
 macroturbulence velocity and instrumental broadening velocity 
 on the basis of \citet{Takeda2008}.
 We used the spectrum of the Sun in \citet{Kuruxz1984} to calculate $v \sin i$ of the Sun.
 \\ \\
 According to \citet{Takeda2008}, a simple relation holds among the line-broadening parameters, which can be expressed as:
 
 \begin{equation}
 v_{\mathrm{M}}^{2} = v_{\mathrm{ip}}^{2}+v_{\mathrm{rt}}^{2}+v_{\mathrm{mt}}^{2}.
 \end{equation}
 
 \noindent $v_{\mathrm{M}}$ is $\mathit{e}$-folding width of the Gaussian macrobroadening 
 function [$f(v)\propto \exp(-(v/v_{\mathrm{M}})^{2})$] including instrumental 
 broadening ($v_{\mathrm{ip}}$), rotation ($v_{\mathrm{rt}}$), and macroturbulence ($v_{\mathrm{mt}}$). 
 We derived $v_{\mathrm{M}}$ by applying automatic spectrum-fitting technique, given the model atmosphere corresponding 
 to the atmospheric 
 parameters (e.g, \cite{Takeda1995}; \cite{Takeda2007}; Takeda et al. 2008). In this process, we used the MPFIT program contained 
 in SPTOOL software package. 
 Although \citet{Takeda2008} applied this fitting technique to the 6080-6089\AA ~ region, we 
 applied this technique to the 
 6212-6220\AA ~ region, which contains four Fe I lines. This is because the 6080-6089\AA ~ region is out of the 
 spectral coverage of our observation.
 In deriving $v_{\mathrm{M}}$ of KIC6934317, 59Vir, and 61Vir, we used the atmospheric 
 parameters derived in Section 3.1 (See also Table 2 and 3).
 In order to calculate $v_{\mathrm{M}}$ of 
 the Sun, we used the values of atmospheric parameters in \citet{Kurucz1993} 
 ($\log g=4.44$, $T_{\mathrm{eff}}=5777$K). We also determined 
 $v_{\mathrm{t}}$ of the Sun is 1 km $\mathrm{s}^{-1}$.
 \\ \\
 $v_{ip}$ is e-folding width of the Gaussian 
 instrumental broadening function. $v_{\mathrm{ip}}$ was calculated by using the following relation (\cite{Takeda2008}),
 
 \begin{equation}
 v_{\mathrm{ip}} = \frac{3\times 10^{5}}{2R\sqrt{ln 2}},
 \end{equation}
 
 \noindent where $R$ is a resolving power of the observation. $v_{\mathrm{mt}}$ is $\mathit{e}$-folding width of the Gaussian macroturbulence broadening 
 function. $v_{\mathrm{mt}}$ was estimated
 by using the relation, $v_{\mathrm{mt}}\sim 0.42{\zeta}_{\mathrm{RT}}$ (\citet{Takeda2008}). ${\zeta}_{\mathrm{RT}}$ is the radial 
 tangential macroturbulence and we calculated ${\zeta}_{\mathrm{RT}}$ by the relation (\cite{Valenti2005}), 
 
 \begin{equation}
 {\zeta}_{\mathrm{RT}} = \left( 3.98 - \frac{T_{\mathrm{eff}} - 5770K}{650K} \right),
 \end{equation}
 
 \noindent where $T_{\mathrm{eff}}$ is the effective temperature of stars.
 \\ \\
 \noindent Using these equations, we derived $v_{\mathrm{rt}}$, which is $\mathit{e}$-folding width of the Gaussian rotational 
 broadening function, and finally we could derive $v \sin i$ by using the relation $v_{\mathrm{rt}}\sim 0.94 v \sin i$ 
 between $v_{\mathrm{rt}}$ and $v \sin i$ (\cite{Gray2005}). 
 \citet{Hirano2012} estimated the systematic uncertainty for $v \sin i$ by chaging ${\zeta}_{\mathrm{RT}}$ by
 $\pm$15\% from Equation (4) for cool stars ($T_{\mathrm{eff}} \leq $6100K) on the basis of observed 
 distribution of ${\zeta}_{\mathrm{RT}}$ (See also Figure 3 in \cite{Valenti2005}). They explained that 
 the statistical errors in fitting each spectrum are generally smaller than the systematic errors arising from different values 
 of ${\zeta}_{\mathrm{RT}}$.
 \\ \\

\subsection{Measurements of Ca II Infrared Triplet (Ca II IRT) and H$\alpha$}

\noindent In order to investigate the chromospheric activity of the target stars, we used $r_{0}(8498)$, $r_{0}(8542)$, and $r_{0}(8662)$, which are 
 the residual flux normalized by the continuum at the line cores of Ca II Infrared Triplet 
 (Ca II IRT, $\lambda$ =8498, 8542, 8662\AA). As the chromospheric activity is enhanced and plages 
 appear, $r_{0}(8498)$, $r_{0}(8542)$, and $r_{0}(8662)$ become large, since a greater 
amount of emission from the chromosphere fills in the core of the lines 
(e.g., \cite{Linsky1979}; \cite{Foing1989}; \cite{Takeda2010}). These indicators are known to 
correlate well with the Ca II H+K emission index (log$R_{\mathrm{HK}}'$), which has been used as the indicator of chromospheric 
activity more traditionally (e.g., \cite{Rutten1984}; \cite{Schrijver1992}; \cite{Chmielewski2000}; \cite{Busa2007}; 
\cite{Hall2008}; \cite{Takeda2012}). We also used $r_{0}$(H$\alpha$) as an indicator of chromospheric 
activity. \citet{Herbig1985}, \citet{Soderblom1993}, and \citet{Frasca2010} described
that H$\alpha$ is also a useful indicator of chromospheric activity, 
and $r_{0}$(H$\alpha$) is correlated well with Ca II H+K emission index (log$R_{\mathrm{HK}}'$). 
On the other hand, Ca II IRT lines are more easily used as diagnostics of chromospheric activity compared to Ca II H\&K 
lines, because  they lie in a wavelength region where the continuum is well defined and the spectrograph detectors 
are more efficient (\cite{Frasca2010}).
\\ \\
We also used emission flux of Ca II IRT and H$\alpha$ lines. The residual flux at the cores of 
Ca II IRT and H$\alpha$ lines, as defined in the previous paragraph, is an indicator of chromspheric activity which is widely 
used (e.g., \cite{Linsky1979}; \cite{Foing1989}), but it also depends on the value of $v \sin i$ of the star (e.g., \cite{Takeda2010}).
A large value of $v \sin i$ can indeed rise the residual flux, mimicking the effect of filling the 
line core with chromospheric emission. 
In order to remove this influence of $v \sin i$, we used the spectral subtraction technique
(e.g., \cite{Frasca1994}; \cite{Frasca2011}; \cite{Martinez-Arnaiz2011}; \cite{Frohlich2012}). 
\citet{Martinez-Arnaiz2011} pointed out that this subtraction process also permits the subtraction of the underlying photospheric 
contribution from the spectrum of the star, 
and in this way we could investigate the spectral emission originated from the chromosphere in detail.
We used the spectrum of 61Vir obtained in this observation as an inactive template to be subtracted from 
the spectrum of KIC6934317. 61Vir is a slowly rotating and non-active early 
G-type main sequence star (e.g., \cite{Anderson2010}), whose atmospheric 
parameters are very similar to those of KIC6934317 (See Tables 2 and 3).
\\ \\
We measured the excess equivalent width ($W^{\mathrm{em}}_{\lambda}$) of the Ca II IRT and H$\alpha$ lines 
in the residual spectrum resulting from this subtraction process.
We subsequently derived emission 
fluxes ($F^{\mathrm{em}}_{\lambda}$) of Ca II IRT and H$\alpha$ lines from $W^{\mathrm{em}}_{\lambda}$ of these lines by 
the following relation ,
$F^{\mathrm{em}}_{\lambda}=W^{\mathrm{em}}_{\lambda}F^{\mathrm{cont}}_{\lambda}$ (\citet{Martinez-Arnaiz2011}).
$F^{\mathrm{cont}}_{\lambda}$ is the continuum flux around the wavelength of each line. 
We calculated $F^{\mathrm{cont}}_{\lambda}$ by using the the following empirical 
relationships between $F^{\mathrm{cont}}_{\lambda}$ and color index ($B-V$) derived by \citet{Hall1996},

 \begin{equation}
 \log F^{\mathrm{cont}}_{\mathrm{H}\alpha} = [7.538 - 1.081(B-V)]\pm 0.033, \ +0.0\leq B-V \leq +1.4,
 \end{equation}

 \begin{equation}
 \log F^{\mathrm{cont}}_{\mathrm{IRT}} = [7.223 - 1.330(B-V)]\pm 0.043, \ -0.1\leq B-V \leq +0.22,
 \end{equation}
 
 \begin{equation}
 \log F^{\mathrm{cont}}_{\mathrm{IRT}} = [7.083 - 0.685(B-V)]\pm 0.055, \ +0.22\leq B-V \leq +1.4.
 \end{equation}

\noindent \citet{Martinez-Arnaiz2011} also used the same method to calculate $F^{\mathrm{cont}}_{\lambda}$.
\\ \\
In Section 3.1, we show that KIC6934317 is $T_{\mathrm{eff}}=5694\pm 25$K and [Fe/H]$=-0.03\pm 0.07$.
\citet{Alonso1996} showed the empirical relations of low mass main sequence stars (F0V$-$K5V) among [Fe/H], $T_{\mathrm{eff}}$, 
and color index. 
We estimated the unreddened color $(B-V)_{0}$ of this star as $\sim$ 0.64 by using this relation. 
Photometric data of this star (\cite{Everett2012}) show that KIC6934317 is $13.191\pm 0.023$ mag in the B-band and 
$12.516\pm 0.017$ mag in the V-band (See Table 1), and thus $B-V$ simply derived from these photometric data is $0.675\pm 0.040$. 
The KIC (\cite{Brown2011}) reports a value of color excess E($B-V$)=$0.093\pm 0.1$ mag, 
which has been estimated with a simple model for the dust distribution in the Milky Way.
\\ \\
The value of $B-V$ estimated from atmospheric parameters derived spectroscopically ($(B-V)_{0}$)
agrees with the corrected value of $B-V$ on the basis of the photometric data within the error range, and is considered to 
be more accurate than the corrected value on the basis of photometric data.
We consequently use the value estimated from atmospheric parameters derived spectroscopically ($(B-V)_{0}$)
and the resulting color excess E($B-V$) $\sim 0.04\pm 0.04$ mag in the following discussions.
\\ \\
\citet{Martinez-Arnaiz2011} described that emission fluxes ($F^{\mathrm{em}}_{\lambda}$) of Ca II IRT and H$\alpha$ 
lines are correlated well with one another, and are also correlated well with $F^{\mathrm{em}}_{\lambda}$ of Ca II H \& K lines, 
which have been used as the indicator of chromospheric activity more traditionally (e.g., \cite{Rutten1984}; \cite{Schrijver1992}; 
\cite{Chmielewski2000}; \cite{Busa2007}; \cite{Hall2008}; \cite{Takeda2012}). Furthermore, $F^{\mathrm{em}}_{\lambda}$ of 
Ca II IRT, Ca II H \& K, and H$\alpha$ lines are correlated well 
with X-ray surface flux ($F_{\mathrm{X}}$), which indicates coronal activity (e.g., \cite{Schrijver1992}; \cite{Martinez-Arnaiz2011}).
\\ \\

\section{Results}
\label{sec:Results}

\subsection{Projected Rotational Velocity}
\noindent Figure 3 shows that the photospheric lines of 59Vir are shallow and broad, indicating 59Vir is a fast rotator.
On the other hand, the line of KIC6934317 is narrow and deep like those of 61Vir and the Sun, which rotate slowly. 
Using the methods explained in Section 3.2, the values of $v \sin i$ of KIC6934317, 59Vir, 61Vir and the Sun 
are about 1.91, 6.27, 1.38, and 1.82 km $\mathrm{s}^{-1}$, respectively.
We show these results in Table 2 and 3. 
The results of 59Vir and 61Vir of this observation are roughly consistent with those of previous researches which are
described in Table 3 (e.g., \cite{Anderson2010}, \cite{Schroder2009}).
In Section 5.2, we discuss the results of $v \sin i$ in detail.
\\ \\

\subsection{Radial Velocity}
\noindent We measured the radial velocity (RV) of the target stars 
by using about 65 Fe I lines. As we showed 
in Table 2 and 3, the RV values of KIC6934317, 59Vir, and 61Vir measured on 
our spectra are -12.238$\pm$0.018, -26.727$\pm$0.026, and -7.528$\pm$0.027 km $\mathrm{s}^{-1}$, respectively. 
These values are derived from the spectra obtained by integrating all frames of spectra of individual stars.
The errors of RV values are the standard errors of the mean RV values which are calculated by using the 
values derived from individual lines.
The results for 59Vir and 61Vir agree with those of previous studies (e.g., \cite{Takeda2005a}), which are described in Table 3.
In Section 5.3, we discuss the results of RV in detail. 
\\ \\

\subsection{Stellar Age, Mass, Luminosity, Radius, and Distance}
\noindent In Section 3.1, we derived $T_{\mathrm{eff}}$, $\log g$, and [Fe/H] of target stars from our observed spectra. 
On the basis of these parameters, we can estimate the stellar age, the stellar mass ($M_{\mathrm{s}}$), the absolute V magnitude ($M_{\mathrm{V}}$), and the stellar 
bolometric luminosity ($L_{\mathrm{s}}$) for the target stars by applying the isochrones of \citet{Girardi2000}. In addition, using the values of 
$L_{\mathrm{s}}$ and $T_{\mathrm{eff}}$, we can also estimate the stellar radius ($R_{\mathrm{s}}$).
\\ \\
The values of the stellar age, $M_{\mathrm{s}}$, $M_{\mathrm{V}}$, $L_{\mathrm{s}}$, and $R_{\mathrm{s}}$ of KIC6934317
are  $\sim 4.5-7.0$ Gyr, $\sim 0.97$\MO, $\sim 4.9-5.0$ mag, $\sim 0.9$\LO, and $\sim 1.0$\RO, respectively. 
Though the range of stellar age is wide, this value is roughly similar to the age of the Sun.
\\ \\
In addition, Table 1 shows that the apparent V magnitude of this star is $12.516\pm 0.017$ mag.
Using the value of E($B-V$) estimated in Section 3.3, the value of interstellar extinction 
in the V band (A(V)) is A(V)$=3.1\times $E($B-V$)$ \sim 0.12\pm0.12$ mag (\cite{Cardelli1989}).
Using these values and $M_{\mathrm{V}}$, we 
estimate that the distance of this star ($d$) is $\sim$ 300 pc.
\\ \\

\subsection{Spectra of Ca II Infrared Triplet and H$\alpha$}
\noindent In the top of Figure 4(a) and 4(b),
normalized spectra of KIC6934317, 59Vir and 61Vir around the cores of 
$\lambda$8498 and $\lambda$8542 \AA ~ are represented, respectively. 
In the bottom of these figures,
the spectrum obtained after the subtraction of the inactive template (61 Vir) is represented.
The excess emissions in the cores of Ca II 8498 \AA~ and 8542 \AA~ are evident in these subtracted spectra.
\\ \\
Using these spectra, we found that $r_{0}(8498)$ and $r_{0}(8542)$ of KIC6934317 
are larger than those of 61Vir, which rotates slowly and has very weak magnetic field, 
while they are comparable to those of 59Vir, which rotates fast ($\sim$6.28 km $\mathrm{s}^{-1}$) and has strong magnetic fields. We show $r_{0}(8498)$, 
$r_{0}(8542)$, $r_{0}(8662)$, and $r_{0}$(H$\alpha$) of KIC6934317, 59Vir, 61Vir, and the Sun in Table 4.
The spectrum of the Sun is obtained by \citet{Kuruxz1984}. 
The original resolving power (R=$\lambda$/$\Delta$$\lambda$) of this solar spectrum is $\sim$500,000, 
and we used the corrected solar spectrum which 
we rebinned from R$\sim$500,000 to R$\sim$100,000
to match the resolution of our HDS spectra.
This table shows that the difference of $r_{0}(8498)$ between KIC6934317 and the Sun is 0.14 and the difference 
of $r_{0}(8542)$ between KIC6934317 and the Sun is 0.15. 
This table shows that the tendencies of $r_{0}(8662)$ and $r_{0}$(H$\alpha$) are similar to those 
of $r_{0}(8498)$ and $r_{0}(8542)$.
\citet{Chmielewski2000} reported a value of $r_{0}(8542) =$ 0.24 for 61Vir. Comparing this data with the results described 
in Table 4, we find that the error range of $r_{0}(8542)$ is $\sim$0.05.
Previous studies (e.g., \cite{Linsky1979}; \cite{Herbig1985}; 
\cite{Chmielewski2000}) have shown that the values of $r_{0}(8498)$, $r_{0}(8542)$, $r_{0}(8662)$, and $r_{0}$(H$\alpha$) become large 
as the star shows high chromospheric activity and large plages appear on the star. Altogether, these results suggest that KIC6934317 
shows high chromospheric activity like 59Vir. 
\\ \\
Table 5 shows the excess equivalent width ($W^{\mathrm{em}}_{\lambda}$) of the Ca II IRT and 
H$\alpha$ lines, and emission flux ($F^{\mathrm{em}}_{\lambda}$) of these lines.
The values of $F^{\mathrm{em}}_{\lambda}$ are derived by using the method explained in Section 3.3.
In Table 5, we report these values not only of KIC6934317, but also of the stars investigated in \citet{Frohlich2012} 
(KIC7985370 and KIC7765135), with the aim of comparing chromospheric activity.
\citet{Frohlich2012} reported that KIC7985370 and KIC7765135 are early G-type main sequence stars. 
They also reported that these stars are young (Age; $100-200$ Myr) and show high chromospheric activity. 
On the basis of the values in Table 5, $F^{\mathrm{em}}_{\mathrm{H}\alpha}$ of KIC6934317 
is roughly as large as of those of KIC7985370 and KIC7765135. In contrast, $F^{\mathrm{em}}_{\lambda}$ of Ca II IRT lines of 
KIC6934317 are about a few times smaller than those of KIC7985370 and KIC7765135.
Consequently, the chromospheric activity of KIC6934317 is fairly high, and this star 
displays an excess of H$\alpha$ flux with respect to Ca II IRT flux, compared to KIC7985370 and KIC7765135. 
\\ \\
In Section 5.1, we further discuss the stellar activity of KIC6934317.
\\ \\

\subsection{Spectra of Li I}
\noindent Figure 5 shows normalized spectra of KIC6934317, 59Vir, 61Vir, and the Sun around the Li I line (6708\AA). 
This figure shows that 59Vir has a deep line of Li, and that the Li line of 61Vir is absent while 
that of the Sun and KIC6934317 are very weak and just visible against the noise.
\\ \\
We measured the lithium abundance (A(Li)\footnote[8]{ A(Li) is defined by the following relation,

 \begin{equation}
 \textnormal{A(Li)} = \log (\mathrm{N}_{\mathrm{Li}}/\mathrm{N}_{\mathrm{H}}) + 12.00,
 \end{equation}
  
\noindent where $(\mathrm{N}_{\mathrm{Li}}/\mathrm{N}_{\mathrm{H}})_{\mathrm{star}}$ is the ratio of the 
number of Lithium (Li) atoms to that of Hydrogen (H) atoms in the star.}) of KIC6934317, 59Vir, 61Vir, and the Sun 
on the basis of the automatic profile 
fitting method described in details in \citet{KTakeda2005}.
In this process of calculating A(Li), we used the MPFIT program contained in SPTOOL software package, and assumed 
LTE (Local Thermodynamical Equilibrium) for the formation of all lines including the Li I line.
We also assumed ${}^{6}\mathrm{Li}/{}^{7}\mathrm{Li}=0$ throughout this study. 
The line lists we used are the same of \citet{KTakeda2005}.
\\ \\
The result of KIC6934317 is A(Li)$=1.25$. In addition, the result of 59Vir is A(Li)$=2.81$ and that of the Sun is A(Li)$=0.88$. 
We could not derive A(Li) of 61Vir, because the Li line of 61Vir is absent.
\\ \\
\citet{KTakeda2005} shows that in the case of calculating A(Li) by LTE, the results 
of 59Vir are $W_{\mathrm{Li}}$\footnote[9]{ $W_{\mathrm{Li}}$ is the equivalent width
of the line of Li I 6708\AA.}$=83.1$ m\AA \ and A(Li)$=2.91$, those of the Sun 
are $W_{\mathrm{Li}}=2.1$ m\AA \ and A(Li)$=0.85$, and those of 61Vir are $W_{\mathrm{Li}}<2$ m\AA \ and A(Li)$<0.86$.
In the case of calculating A(Li) by NLTE (Non Local Thermodynamical Equilibrium), A(Li) of 59Vir and the Sun are 2.91 and 0.92, respectively.
\citet{Asplund2009} also shows that A(Li) of the solar photosphere calculated by NLTE is $1.05\pm 0.10$.
In table 2 and 3, we show the values of previous studies and this study.
The values of 59Vir, 61Vir, and the Sun of this observation are roughly consistent with those of previous studies.
These results show that the lithium abundance of KIC6934317 is quite lower than that of 59Vir, and a bit higher than that of 
61Vir and the Sun. 
\\ \\
It has been known that the lithium abundances could provide a fairly rough constraint on the age of early 
G-type stars (e.g., \cite{Soderblom1993a}; \cite{Sestito2005}), 
though there are a lot of unknown problems about the behavior of lithium abundances in early 
G-type stars (e.g., \cite{Ryan2001}; \cite{Israelian2004}; \cite{Sestito2005}; \citet{Takeda2007b}; 
\cite{Takeda2010}; \cite{Takeda2012}).
On the basis of the fairly rough relation between A(Li) and the stellar age discussed in \citet{Sestito2005},
\citet{Takeda2007b}, and \citet{Takeda2010}, the age of KIC6934317 is more than about a few Gyr. 
This value is roughly consistent with the age derived by using the isochrones of \citet{Girardi2000} 
(see Section 4.3), though the ranges of these two values are wide.
\\ \\

\section{Discussion} 
\label{sec:Discussion}    

\subsection{Stellar Activity and Superflares}
\noindent In Figure 6, we plot $r_{0}(8542)$ as a function of effective temperature ($T_{\mathrm{eff}}$) of the stars.  
In making this figure, we use both the results described in Table 4 and the data from \citet{Chmielewski2000}. 
There appears to be a clear dividing gap between active and quiescent stars which show some slope with effective 
temperature (\cite{Foing1989}, \cite{Chmielewski2000}). 
$r_{0}(8542)$ become large as the activity of 
chromosphere is enhanced and plages appear (e.g., \cite{Linsky1979}; \cite{Foing1989}; \cite{Chmielewski2000}; \cite{Takeda2010}). 
Nearly all giant and sub-giant stars ($\log g$$\leq$4.0) are not active, unless
they belong to close binary systems. This is because they are old and the magnetic braking during their main-sequence lifetime and 
the star expansion in the red giant phase spin down the star. As a consequence they display a low 
chromospheric activity (\cite{Takeda2010}). 
\\ \\
In addition, this clear dividing gap between active and quiescent stars is considered to 
correspond to Vaughan$-$Preston gap in Ca II H\&K (\cite{Vaughan(1980)}).
\citet{Martinez-Arnaiz2011} reported that the Vaughan$-$Preston gap is also dividing 
in two separate classes stars with large emission flux
and those with small emission flux in other chromospheric lines like H$\alpha$.
It is widely accepted that the differences of stellar age and dynamo regimes are deeply related to 
exisitence of this gap between active and quiescent stars (e.g., \cite{Durney1981}; \cite{Noyes1984}; \cite{Bohm-Vitense2007};
\cite{Pace2009}; \cite{Martinez-Arnaiz2011}). It is also believed that nanoflare 
heating, which is supposed to be one of the main heating 
mechanism of the stellar outer atmospheres, contributes to form the gap (\cite{Martinez-Arnaiz2011}). 
\\ \\
As apparent from Figure 6, KIC6934317 turns out to be a relatively active star, in spite of  
its low lithium abundance (see Section 4.5). This figures also indicates 
that 59Vir is an active star, and 61Vir and the Sun are not active stars. These results of 59Vir, 61Vir, and the Sun are 
consistent with the findings of previous researches (e.g. \cite{Anderson2010}, \cite{Takeda2012}). 
\\ \\
\citet{Martinez-Arnaiz2011} plotted the emission flux in H$\alpha$ versus $B-V$, and displayed the existence of 
the Vaughan$-$Preston gap. 
On the basis of that relation, G-type stars above the gap (active stars) have
$\log F^{\mathrm{em}}_{\mathrm{H}\alpha} \sim 6-7$. 
With a $\log F^{\mathrm{em}}_{\mathrm{H}\alpha}=6.3$ and $(B-V)_{0} \sim 0.64$ (See Section 3.3 and Table 5), 
KIC6934317 is located above the gap, among the very active stars where strong flares are frequent.
\\ \\
The comparison between our results of emission flux in Table 5 (KIC6934317 has an excess H$\alpha$ flux as compared to Ca II IRT fluxes.) 
and the results in \citet{Martinez-Arnaiz2011} (A star above the Vaughan$-$Preston gap has an excess of 
H$\alpha$ flux with respect to Ca II IRT fluxes.) also shows that this star belongs to the active star group above the gap.
\\ \\
X-ray flux is widely used as a measure of coronal activity and a direct measure of stellar magnetic activity,
because it is unlikely to include contribution from other sources such as basal atmosphere (e.g., \cite{Schrijver1992}; 
\cite{Pevtsov2003}; \cite{Pizzolato2003}). 
\citet{Martinez-Arnaiz2011} presented the emprical relationship between H$\alpha$ flux and 
X-ray flux ($F_{\mathrm{X}}$), and this empirical relationship is given by the following equation,

\begin{equation}
\log F_{\mathrm{X}} = (-2.98\pm 0.39)+(1.60\pm 0.07)\log F^{\mathrm{em}}_{\mathrm{H}\alpha}.
\end{equation}

\noindent Using this equation, we estimate that $F_{\mathrm{X}}$ of KIC6934317 is 
$F_{\mathrm{X}} \sim 1 \times 10^{7}$ erg $\mathrm{cm}^{-2}$$\mathrm{s}^{-1}$, 
the X-ray luminosity ($L_{\mathrm{X}}$) of this star is 
$L_{\mathrm{X}}\sim 4\pi R_{\mathrm{s}}^{2} F_{\mathrm{X}} \sim 8 \times 10^{29}$ erg $\mathrm{s}^{-1}$,
and $\log L_{\mathrm{X}}$/$L_{\mathrm{s}}$ is $\sim -3.6$. In calculating these values, we used the values of
$R_{\mathrm{s}} \sim 1.0\RO$ and $L_{\mathrm{s}} \sim$ 0.9\LO, as calculated in Section 3.3.
\\ \\
$L_{\mathrm{X}}$ estimated for KIC6934317 is as large as that of KIC7985370 (an active early G-type main sequence star), which is 
$L_{\mathrm{X}}= (3-7)\times 10^{29}$ erg $\mathrm{s}^{-1}$ (\cite{Frohlich2012}), and 
much larger than that of the Sun, which varies from about $5\times 10^{26}$ erg $\mathrm{s}^{-1}$ to 
about $2\times 10^{27}$ erg $\mathrm{s}^{-1}$ from the minimum to the maximum of the 
activity cycle (e.g., \cite{Schmitt1995}).
Hence KIC6934317 is considered to have high coronal activity.
\\ \\
Nevertheless, we do not find the X-ray counterpart of this star in the ROSAT All-Sky Survey (RASS) Faint 
Source Catalog (\cite{Voges2000}). 
Indeed, the detection limit of the X-ray flux for the RASS survey
is $\sim 2\times 10^{-13}$ erg $\mathrm{cm}^{-2}$$\mathrm{s}^{-1}$ (\cite{Schmitt1995}), 
which corresponds to the X-ray luminosity of $\sim 2\times 10^{30}$ erg $\mathrm{s}^{-1}$ assuming the distance of $\sim$ 300 pc.
This explains why KIC6934317 was not detected by the RASS.
\\ \\
\noindent On the basis of the discussions and considerations described in this Section and the results of observations 
that we mentioned in Section 4.3, KIC6934317 shows high chromospheric and coronal activity 
and it very likely has large plages.
In addition, a star with such a high chromospheric activity is also believed
to have large starspots 
(e.g., \cite{Schrijver1992}; \cite{Busa2007}; \cite{Hall2008}). Therefore, given its high activity level, we believe
that KIC6934317 has large starspots which can store a large amount of magnetic energy that 
cause superflares (\cite{Shibata2013}).
\\ \\

\subsection{Rotational Velocity and Inclination Angle}
\noindent The value of $v \sin i$ of KIC6934317 estimated by this spectroscopic 
observation is $\sim$1.91 km $\mathrm{s}^{-1}$, 
$R_{\mathrm{s}}$ estimated in Section 3.3 is $\sim $1.0\RO, and 
$P_{\mathrm{s}}$ is about 2.54 days. Assuming 
that the brightness variation of this star is caused by the 
rotation of the star with starspots, we estimate 
rotational velocity (\textit{v}) as $\sim$20 km $\mathrm{s}^{-1}$ by using following relation:

\begin{equation}
 \textit{v} = \frac{2\pi R_{\mathrm{s}}}{P_{\mathrm{s}}}
\end{equation} 

\noindent We think that the difference between the values of \textit{v} and $v \sin i$ can be 
explained by the inclination effect. Once \textit{v} is estimated, 
we estimate that \textit{i} (the stellar inclination angle) is $\sim$5.5 deg by the following relation (\cite{Frasca2011}, \cite{Hirano2012}):

\begin{equation}
 \textit{i} = \arcsin{\left[ \frac{(v \sin i)_{\mathrm{spec}}}{\textit{v}} \right]} 
\end{equation} 

\noindent This result suggest that KIC6934317 has a small value of \textit{i} and is a nearly pole-on star.
\\ \\
\noindent We can confirm this inclination effect from another point of view. Figure 7 shows the scatter plot of the superflare 
amplitude as a function of the 
amplitude of the brightness variation. The data are taken from \citet{YNotsu2013}. The solid line corresponds to the analytic relation 
between the stellar brightness variation amplitude and superflare amplitude for $B$=1000G (\cite{YNotsu2013}). The amplitude 
of the brightness variation of the superflare star is considered to be a good indicator of the starspot coverage. In addition, the energy
of the superflare can be estimated with the superflare amplitude and duration time.
The dashed line corresponds to the same relation in case of nearly pole-on 
(\textit{i}=2.0 deg) for $B$=1000G, assuming that the brightness variation of a star become small as the
inclination angle of the star 
become small. These lines are considered to give an upper limit for the flare amplitude in each inclination angle. These results 
suggest that \textit{i} of KIC6934317 is small (nearly pole-on) and that this star has large starspots 
generating superflares, though the stellar brightness variation amplitude of this star is small ($\sim 0.1 \% $).
\\ \\

\subsection{Binarity}
\noindent On the basis of the stellar parameters derived in this study, 
this star proves to be an early G-type main sequence 
star.
\citet{Noyes1984} argued that the rotational period is correlated with the chromospheric activity, which is known to be 
an indicator of the magnetic activity of the star. \citet{Pallavicini1981} and \citet{Pizzolato2003} 
showed that rapidly rotating stars have high coronal and magnetic activity than slowly rotating stars.
On the basis of these studies, 
it is not strange that KIC6934317 show high chromospheric and magnetic activity. This is because KIC6934317 
has the high rotational velocity (\textit{v} $\sim$20 km $\mathrm{s}^{-1}$).
\\ \\
\citet{Ayres1997} discussed the dependence of 
stellar rotational velocity on age for solar-type stars and presented the following relation,

\begin{equation}
 \frac{\textit{v}}{v_{\mathrm{sun}}} = \left[\frac{t}{t_{\mathrm{sun}}}\right]^{-0.6\pm 0.1},
\end{equation} 

\noindent where \textit{v} is the rotational velocity of a star, $v_{\mathrm{sun}}$ is the rotational velocity of the 
Sun ($\sim$1.82 km $\mathrm{s}^{-1}$), \textit{t} is age of a star, and $t_{\mathrm{sun}}$ is the solar 
age ($\sim$4.6 [Gyr]).
Using this relation and rotational velocity of this star (\textit{v} $\sim$20 km $\mathrm{s}^{-1}$),
we can estimate that the age of KIC6934317 is $\sim$100 Myr. 
This value is incompatible with the value estimated in this observation by using the lithium abundance of this star, and the 
isochrones (See Section 4.3 and 4.5), though there are a lot of unknown problems about the behavior of lithium abundances in 
early G-type stars as we mention in Section 4.5, and the ranges of values estimated by using these two methods are wide.
Thus, one hypothesis that explains the large rotational velocity and high level of activity would be that this star is a binary 
which has maintained a high rotation rate thanks to the tidal interaction that has coupled spin and orbit (\cite{Walter1981}).
\\ \\
First, we checked the slit viewer images of Subaru/HDS for 
this star taken simultaneously with the spectra, and we cannot find companions for this star as far as we looked this images.
\\ \\
Second, if a star is a binary star and has a companion, RV of the primary star is expected to 
change between the observations. We estimate the value of the RV change by using the following relation,

\begin{equation}
  f(M) = \frac{(M_{\mathrm{2}}\sin i)^{3}}{(M_{\mathrm{1}}+M_{\mathrm{2}})^{2}} = \frac{P_{\mathrm{o}}K_{\mathrm{1}}^{3}}{2\pi G},
\end{equation} 

\noindent where G is the gravitational constant (= $6.67\times 10^{-11}$ $\mathrm{m}^{3} \mathrm{s}^{-2} 
\mathrm{kg}^{-1}$), $M_{\mathrm{1}}$ is the 
mass of the primary star, $M_{\mathrm{2}}$ is that of the companion, and $f(M)$ is the mass function. $K_{\mathrm{1}}$ is 
the amplitude of radial 
velocity variation of the primary star, and $P_{\mathrm{o}}$ is the period of orbital motion of the binary system. 
Thanks to the parameters and $v \sin i$ derived from our obtained spectrum, we can reasonably assume \textit{i} $=$ 5.5 deg and 
$M_{\mathrm{1}} \sim 0.97\MO$. In addition, if we suppose that this is a close binary system 
and these stars are expected to be tidally locked with each other, we can assume that $P_{\mathrm{o}}$ is $\sim$ 2.54
days (equal to the rotation period derived from the brightness variation), and that the value of 
$K_{\mathrm{1}}$ is larger than the 
$v \sin i$ ($\sim$1.91 km $\mathrm{s}^{-1}$). Adopting the minimum value of $K_{\mathrm{1}}$, we can derive a minimum mass for the unseen secondary component, 
$M_{\mathrm{2}} \sim 0.14\MO$.
\\ \\
\noindent In this observation, we get 6 frames of spectra of KIC6934317, and the exposure time per frame is 1800 
seconds (see Section 2.2). Since a frame was obtained $\sim$ 0.5 hours after the previous frame, RV of the star is
expected to change between the frames if this star is a binary. Actually, RV of the star calculated from these six frames 
are, in order from the time, -12.311 $\pm$0.034, -12.279 $\pm$0.027, -12.190 $\pm$0.028, -12.224 $\pm$0.030, 
-12.218 $\pm$0.029, -12.187 $\pm$0.026 km $\mathrm{s}^{-1}$
(As we explained in Section 4.2, RV of the star derived from the spectra obtained by integrating all frames of spectra of this star is -12.238$\pm$0.018 km $\mathrm{s}^{-1}$). 
The difference between the values of RV of the first frame and the last frame is 0.124 $\pm$0.080 km $\mathrm{s}^{-1}$,
and thus RV of this star is supposed to change between the first and the last frames.
Considering the minimum value of $K_{\mathrm{1}}$ described in the former paragraph and 
the time interval between the two frames, however, we cannot claim
that this star is definitely a binary. This is because the period of observation is small and because the 
values of RV change are very small.
Especially, the values of RV of last four frames are almost the same values within the ranges of errors. In the future we will 
observe KIC6934317 again and estimate the change of RV between the observations in detail 
to investigate whether this star is a binary or not.
\\ \\

\section{Summary}
\noindent In this paper, we present the results of the analysis of a high-resolution spectrum of 
KIC6934317, a star which displays many strong white-light flares (superflares) in the high-precision Kepler light curve.
We measured the core depth and emission flux of Ca II infrared triplet 
lines and H$\alpha$ line in order to investigate the chromospheric activity and 
the presence of starspots. Our analysis shows that KIC6934317 has high chromospheric activity, in spite of
the low lithium abundance and the small amplitude of the rotational modulation.
Using the empirical relations between flux of chrompspheric lines and X-ray flux, 
it should also have a much higher coronal activity than that of the Sun.
We believe that the low amplitude of the rotational modulation evident in the Kepler light curve 
is due only to foreshortening effects and that this star has actually
large starspots which can store a large amount of magnetic energy that can generate superflares.
We also estimated some stellar parameters, such as $T_{\mathrm{eff}}$, $\log g$, [Fe/H], $v \sin i$, and RV.
We confirmed that this star is an early G-type main sequence star, on the basis of these parameters. 
The value of $v \sin i$ of KIC6934317 is $\sim$\textcolor{red}1.91 km $\mathrm{s}^{-1}$, though \textit{v} is estimated to be 
$\sim$20 km $\mathrm{s}^{-1}$ by using the period of the brightness 
variation as the rotation period, and $R_{\mathrm{s}}$ of this star. This difference
between the values of $v \sin i$ and \textit{v} can be explained by small 
inclination angle (nearly pole-on). The amplitude of the brightness 
variation of superflare stars is considered to correspond to the spot coverage, and it is known to 
correlate with the superflare amplitude, which corresponds to the energy of superflare. The relation
between the superflare amplitude 
and the brightness variation amplitude in this star is also found to be explained by the effect of small inclination angle.
On the basis of the lithium abundance and isochrones, it is implied that the age of this star is more than about 
a few Gyr, though the problem of the strong activity at such a high age remains unsolved.
\\ \\
\bigskip

\noindent This study is based on observational data collected at Subaru Telescope, which is operated by the 
National Astronomical Observatory of Japan.
We are grateful to Dr.Akito Tajitsu and other staffs of the Subaru Telescope for making large contributions
in carrying out our observation.
We further thank Dr.Kazuhiro Sekiguchi (NAOJ) and Dr.Yoichi Takeda (NAOJ) for useful advices.
Kepler was selected as the tenth Discovery mission. 
Funding for this mission is provided by the NASA Science Mission Directorate. 
The Kepler data presented in this paper were obtained from the Multimission Archive at STScI. 
This work was supported by the Grant-in-Aid from the Ministry of Education,
Culture, Sports, Science and Technology of Japan (No. 25287039).
\\ \\


\newpage

\begin{table}
  \caption{{\large Photometric data of KIC6934317}}\label{tab:T1}
  \begin{center}
    \begin{tabular}{lll}
     \hline
      RA\footnotemark[$*$]  &(J2000) & $19^{\mathrm{h}}07^{\mathrm{m}}34^{\mathrm{s}}.92$ \\
      DEC\footnotemark[$*$] &(J2000) & $+42^{\circ}29'38''.33$\\
      \hline
      $U$\footnotemark[$\dagger$] && $13.399\pm 0.020$ [mag] \\
      $B$\footnotemark[$\dagger$] && $13.191\pm 0.023$ [mag] \\
      $V$\footnotemark[$\dagger$] && $12.516\pm 0.017$ [mag] \\
      $J$\footnotemark[$\ddagger$] && $10.915\pm 0.019$ [mag] \\
      $H$\footnotemark[$\ddagger$] && $10.570\pm 0.017$ [mag] \\
      $K_\mathrm{s}$\footnotemark[$\ddagger$] && $10.461\pm 0.012$ [mag] \\
      \hline
      \\ 
      \hline
      \multicolumn{3}{@{}l@{}}{\hbox to 0pt{\parbox{120mm}{
      \footnotesize{\large Notes}
       \par\noindent
       \footnotemark[$*$] {\normalsize \citet{Brown2011}}
       \par\noindent
       \footnotemark[$\dagger$] {\normalsize \citet{Everett2012}}
       \par\noindent
        \footnotemark[$\ddagger$] {\normalsize \citet{Skrutskie2006}}}\hss}}
    \end{tabular}
\end{center}
    \end{table}
   
   \newpage

   \begin{table*}
   \caption{{\large Atmospheric parameters, projected rotational velocity, radial velocity, and lithium abundance of KIC6934317}}\label{tab:T2}
   \begin{center}
    \begin{tabular}{llll}
   \hline
   Parameter & & Value & Reference \\
   \hline
   $T_{\mathrm{eff}}$ & [K] & $5694\pm 25$  &present work \\
                      &     & $5387\pm 200$ &1\\
                      &     & $5710\pm 81$  &2\\
                      &     & $5604\pm 103$ &3\\
   $\textnormal{[Fe/H]}$ & & $-0.03\pm 0.07$ &present work\\
                         & & $-0.78\pm 0.5$ &1\\
   $\log g$ & & $4.42\pm 0.08$ &present work\\
            & & $3.8\pm 0.4$ & 1\\
   $v \sin i$ & [km $\mathrm{s}^{-1}$] &1.91& present work\\
   RV & [km $\mathrm{s}^{-1}$] & $-12.238\pm 0.018$& present work\\
   A(Li)& & 1.25 & present work\\
   \hline
   \\
   \hline
   \multicolumn{4}{@{}l@{}}{\hbox to 0pt{\parbox{90mm}{
   \footnotesize{\large References }
   \par\noindent
   \footnotemark[$ $] {\normalsize (1)\citet{Brown2011}}
   \par\noindent
   \footnotemark[$ $] {\normalsize (2)\citet{Pinsonneault2012} (the SDSS method)}
   \par\noindent
   \footnotemark[$ $] {\normalsize (3)\citet{Pinsonneault2012} (the IRFM method)}}\hss}}
    \end{tabular}
    \end{center}
    \end{table*}
   
 \newpage

   \begin{table*}
   \caption{{\large Atmospheric parameters, projected rotational velocity, radial velocity, and lithium abundance of comparison stars}}\label{tab:T3}
   \begin{center}
    \begin{tabular}{lllllllll}
   \hline
   Name&$T_{\mathrm{eff}}$&[Fe/H]&$\log g$&$v \sin i$&RV&$W_{\mathrm{Li}}$&A(Li)&Reference\\
   &[K]&&&[km $\mathrm{s}^{-1}$]&[km $\mathrm{s}^{-1}$]&[m\AA]&&\\
   \hline
   59Vir&$6009\pm 28$&$0.09\pm 0.06$&$4.15\pm 0.06$&6.27&$-26.727\pm 0.026$& &2.81&present work\\
        &6234&&4.60&6.67&&&&1\\
        &6120&0.21&4.25&&&83.1&2.91&2\\
        &5986&&4.25&4.5&&&&3\\
        &&&&&$-25.9$&&&4\\
   61Vir&$5558\pm 15$&$-0.04\pm 0.06$&$4.50\pm 0.04$&1.38&$-7.528\pm 0.027$&&&present work\\
        &5571&&4.47&0.46&&&&1\\
        &5720&0.11&4.67&&&($\leq 2$)&($\leq 0.86$)&2\\
        &5509&&4.51&0.4&&&&3\\
        &&&&&$-8.1$&&&4\\
   \hline
   \\
   \hline
   \multicolumn{9}{@{}l@{}}{\hbox to 0pt{\parbox{180mm}{
   \footnotesize{\large References }
   \par\noindent
   \footnotemark[$ $] {\normalsize (1)\citet{Anderson2010}; (2)\citet{KTakeda2005}; (3)\citet{Schroder2009}; (4)\citet{Takeda2005a}}}\hss}}
      
    \end{tabular}
    \end{center}
    \end{table*}
    
  \newpage

  \begin{table}

  \caption{{\large $r_{0}$ of Ca II IRT and H$\alpha$}}\label{tab:T4}
  \begin{center}
    \begin{tabular}{lllll}
      \hline
      Name & $r_{0}(8498)$\footnotemark[$*$] & $r_{0}(8542)$\footnotemark[$*$] & $r_{0}(8662)$\footnotemark[$*$] & $r_{0}$(H$\alpha$) \footnotemark[$*$]  \\
      \hline
      KIC6934317 & \ 0.43 & \ 0.33 & \ 0.37 & \ 0.42 \\
      59Vir & \ 0.47 & \ 0.36 & \ 0.37 & \ 0.27 \\
      61Vir & \ 0.28 & \ 0.18 & \ 0.19 & \ 0.19 \\
      Sun & \ 0.30\footnotemark[$\dagger$] & \ 0.19\footnotemark[$\dagger$] & \ 0.19\footnotemark[$\dagger$] & \ 0.18\footnotemark[$\dagger$] \\
      \hline
      \\ 
      \hline
      \multicolumn{5}{@{}l@{}}{\hbox to 0pt{\parbox{120mm}{
      \footnotesize{\large Notes}
       \par\noindent
       \footnotemark[$*$] {\normalsize $r_{0}$ is the residual 
       flux normalized by the continuum at the line core. }
       \par\noindent
       \footnotemark[$\dagger$] {\normalsize The spectrum of the Sun is obtained by \citet{Kuruxz1984}. 
       The original resolving 
       power (R=$\lambda$/$\Delta$$\lambda$) of this solar spectrum is $\sim$500,000, but we have brought it to R$\sim$100,000 
       to match the resolution of our HDS spectra.
       }}\hss}}
       \par\noindent
    \end{tabular}
  \end{center}
\end{table}

\newpage

\begin{table}
  \caption{{\large The excess equivalent width and the emission flux}}\label{tab:T5}
  \begin{center}
    \begin{tabular}{lllllll}
      \hline
      Line & \ \ \ \ $W^{\mathrm{em}}_{\lambda}$ & $F^{\mathrm{em}}_{\lambda}$ & \ \ \ \ $W^{\mathrm{em}}_{\lambda}$ & $F^{\mathrm{em}}_{\lambda}$& \ \ \ \ $W^{\mathrm{em}}_{\lambda}$ & $F^{\mathrm{em}}_{\lambda}$\\
           & \ \ \ \ [\AA] & [erg $\mathrm{cm}^{-2}$$\mathrm{s}^{-1}$] & \ \ \ \ [\AA] & [erg $\mathrm{cm}^{-2}$$\mathrm{s}^{-1}$]& \ \ \ \ [\AA] & [erg $\mathrm{cm}^{-2}$$\mathrm{s}^{-1}$]\\
      \hline
    & \ \ KIC6934317& & \ \ KIC7985370\footnotemark[$*$]& & \ \ KIC7765135\footnotemark[$*$]& \\
    & \ (The target star)& & & & &  \\
      Ca II $\lambda$8498& \ \ \ \ \  0.09 &$3.8 \times 10^{5}$ & \  \ \ \ \ 0.26 &$1.3 \times 10^{6}$ & \ \ \ \ 0.26 &$1.3 \times 10^{6}$ \\
      Ca II $\lambda$8542& \ \ \ \ \  0.14 &$6.2 \times 10^{5}$ & \  \ \ \ \ 0.44 &$2.2 \times 10^{6}$ & \ \ \ \ 0.37 &$1.8 \times 10^{6}$ \\
      Ca II $\lambda$8662& \ \ \ \ \  0.12 &$5.5 \times 10^{5}$ & \  \ \ \ \ 0.34 &$1.7 \times 10^{6}$ & \ \ \ \ 0.33 &$1.6 \times 10^{6}$ \\
      H$\alpha$          & \ \ \ \ \  0.29 &$2.0 \times 10^{6}$ & \  $0.16-0.26$ &$(1.2-2.0) \times 10^{6}$ & \ \ \ \ 0.33 &$2.5 \times 10^{6}$ \\  
      \hline
      \\ 
      \hline
      \multicolumn{7}{@{}l@{}}{\hbox to 0pt{\parbox{180mm}{
      \footnotesize{\large Notes}
      \par\noindent
       \footnotemark[$*$] {\normalsize We use the values of KIC7985370 and KIC7765135 presented in \citet{Frohlich2012}.
       These stars are early G-type main sequence stars. They are also young, and show high chromospheric activity.}}\hss}}
       \par\noindent
    \end{tabular}
  \end{center}
\end{table}
    
  \newpage


\begin{figure}
  \begin{center}
    \FigureFile(150mm,150mm){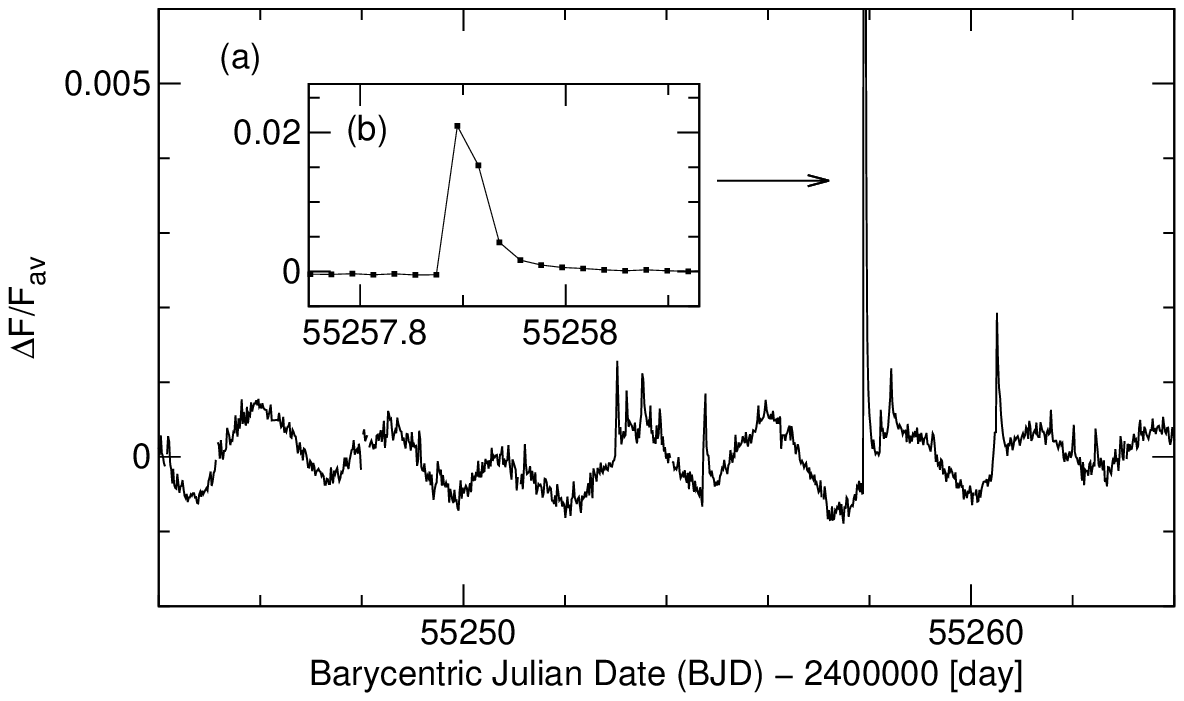}
    \FigureFile(150mm,150mm){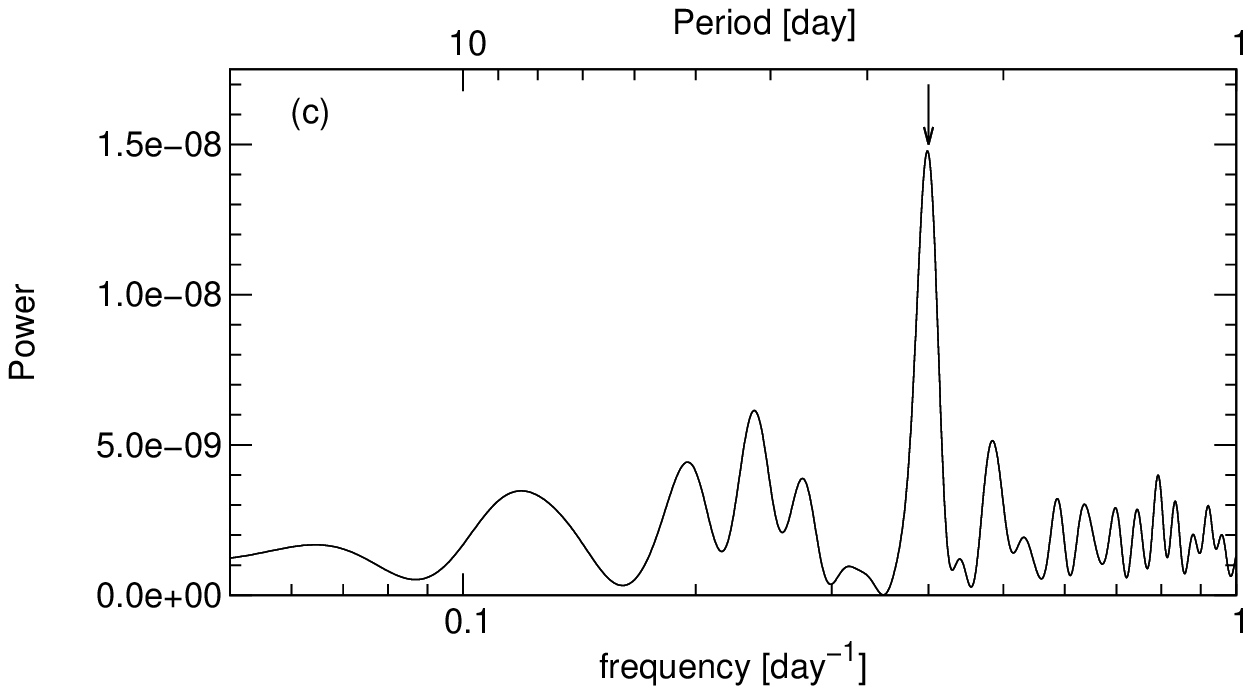}
  \end{center}
  \caption{\normalsize(a): Lightcurve of KIC6934317 obtained by the Kepler spacecraft. The vertical axis is the brightness 
  variation to the average brightness. We can see quasi-periodic modulation with an about 0.1\% amplitude of the total 
  luminosity of the star. (b): The enlarged lightcurve of a superflare observed around BJD 2455257.9. The amplitude of this 
  superflare is about 2.1\% of the brightness of the star. The duration of the flare 
  is about 0.12 day, and the total radiative energy released during this event is estimated to 
  be about $5.6 \times 10^{34}$ erg. (c): Power spectra of the light curves 
  of KIC6934317. This figure indicates that the period of 
  the modulation of KIC6934317 ($P_{\mathrm{s}}$) is 2.54 days.}\label{fig1}
\end{figure}

\newpage

\begin{figure}
  \begin{center}
    \FigureFile(120mm,120mm){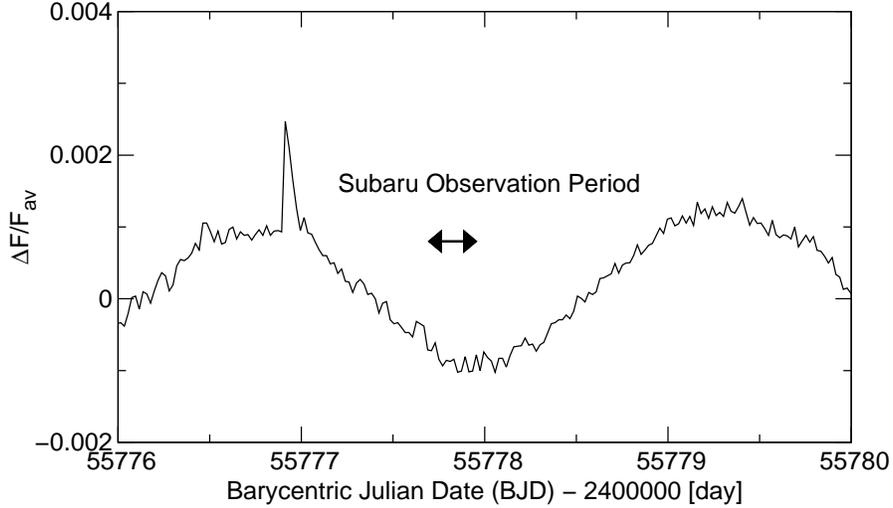}
  \end{center}
  \caption{\normalsize Lightcurve of KIC6934317 around the period we 
  observed this star with Subaru telescope (August 3, 2012 (HST)). The observation period is in the dark phase of brightness variations. 
  There is a flare about a day before the observation period.}\label{fig2}
\end{figure}

\newpage

\begin{figure}
  \begin{center}
    \FigureFile(120mm,120mm){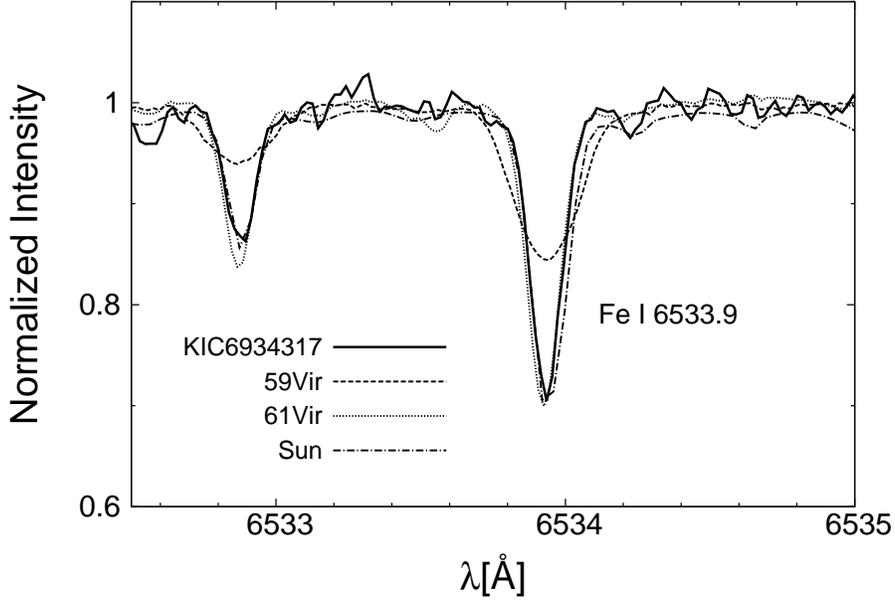}
  \end{center}
  \caption{\normalsize Normalized spectra of KIC6934317 (solid line), 59Vir (dashed line), 61Vir (dotted line), and 
  the Sun (dash-dotted line)around Fe I 6533.9.
  The spectrum of the Sun is obtained by \citet{Kuruxz1984}. 
  The original resolving power (R=$\lambda$/$\Delta$$\lambda$) of this solar spectrum 
  is $\sim$500,000, but we have brought it to R$\sim$100,000 to match the resolution of our HDS spectra.
  These spectra show that 59Vir is a fast rotator, while the other stars and the Sun have 
  a low $v \sin i$.}\label{fig3}
\end{figure}


\begin{figure}
  \begin{center}
    \FigureFile(100mm,100mm){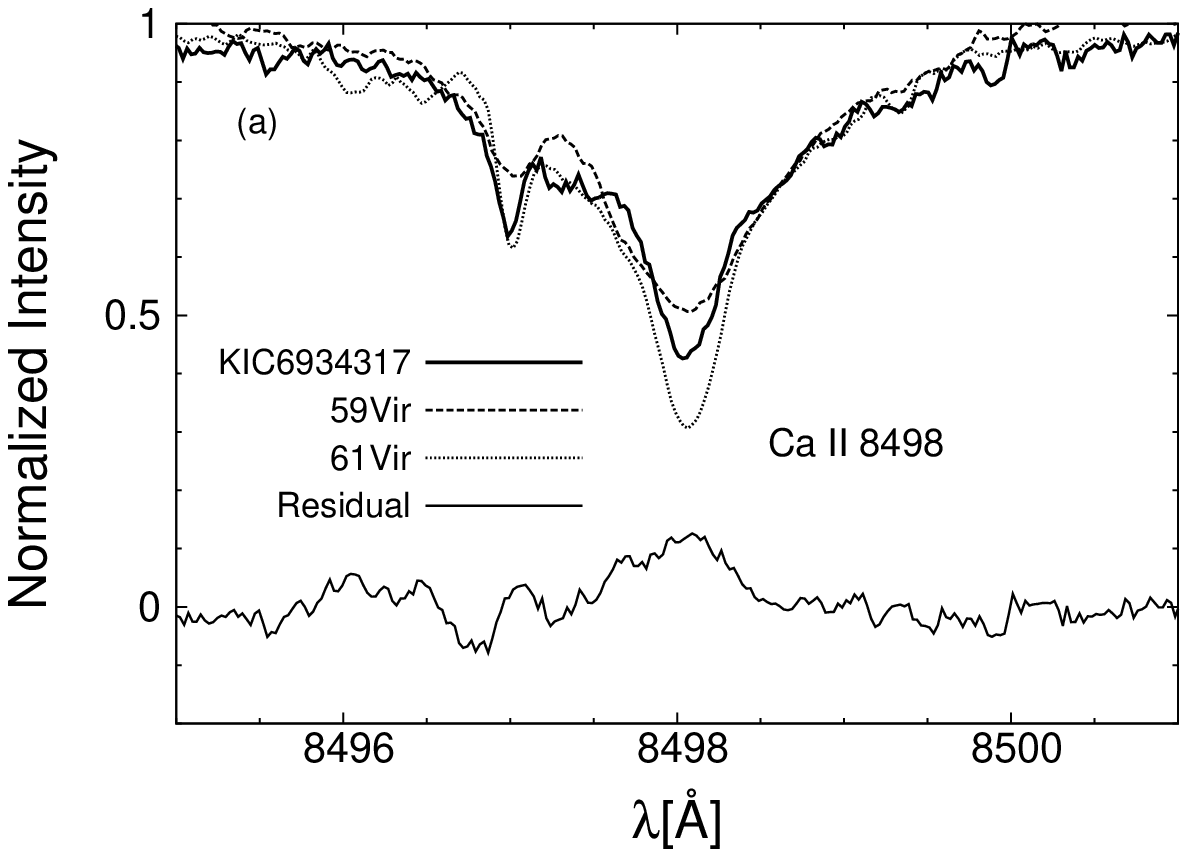}
    \FigureFile(100mm,100mm){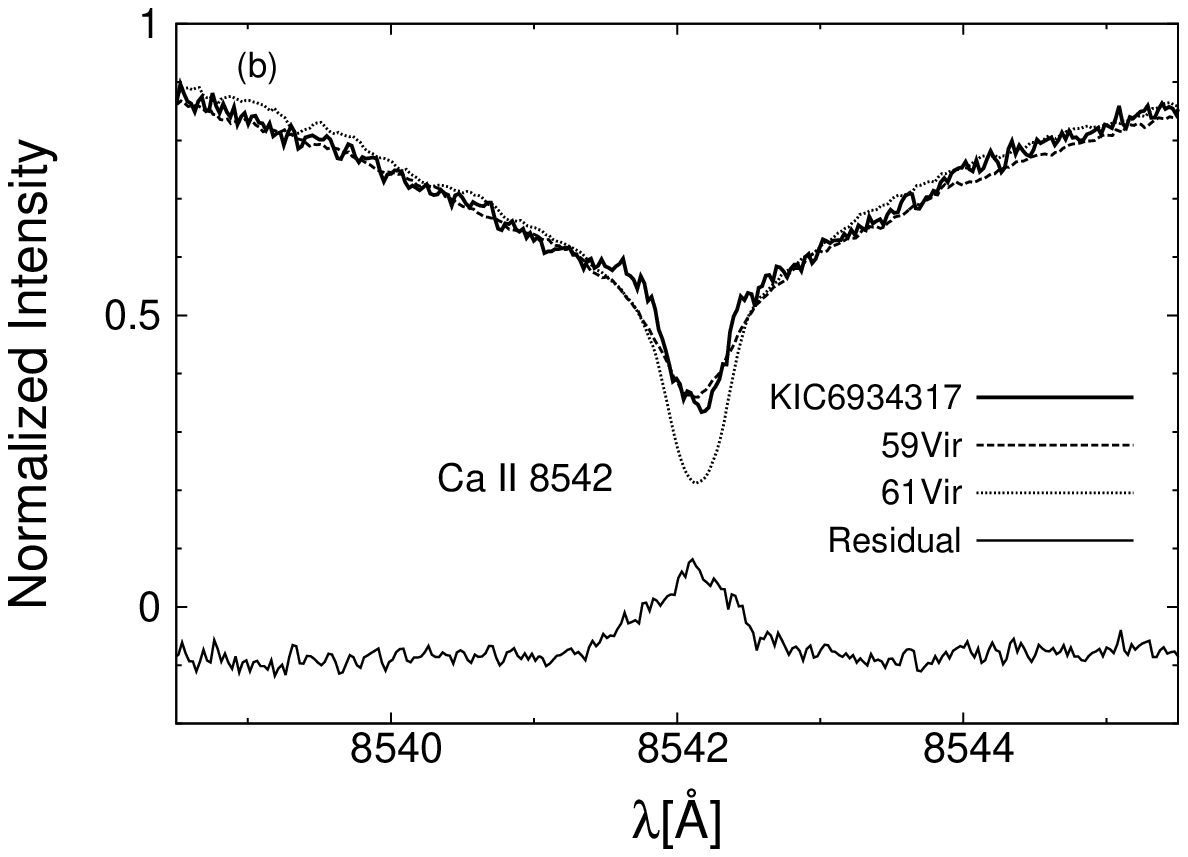}
  \end{center}
  \caption{\normalsize \textit{Top of each panel}: Normalized spectra of KIC6934317 (thick solid line), 59Vir (dashed line), 
  and 61Vir (dotted line) around the core-regions of Ca II 8498 and 8542.  
  59Vir and 61Vir are the comparison stars which have strong and weak magnetic fields, respectively (e.g., \cite{Anderson2010}). 
  These spectra show that $r_{0}(8498)$ and $r_{0}(8542)$ of KIC6934317 are larger than those of 61Vir, and are similar to 
  those of 59Vir. Previous studies (e.g., \cite{Linsky1979}; \citet{Herbig1985}; 
  \cite{Chmielewski2000}) indicate that $r_{0}(8498)$ and $r_{0}(8542)$ become large 
  as the star shows high chromospheric activity. Therefore these results suggest that KIC6934317 shows high chromospheric 
  activity like 59Vir. We show the values of $r_{0}(8498)$ and $r_{0}(8542)$ of these stars in Table 4.
  \textit{Bottom of each panel}: The spectrum after subtracting the spectrum of 
  61Vir (the inactive template star) from KIC6934317 (thin solid line).
  The residual Ca II 8542 profile is plotted shifted downwards by 0.1 for the sake of clarity. There are clear excess
  emission around the core-regions of Ca II 8498 and 8542, and we integrated the excess emission regions to get excess equivalent width 
  and emission flux (See Table 5).}\label{fig4}
\end{figure}

\newpage

\begin{figure}
  \begin{center}
    \FigureFile(120mm,120mm){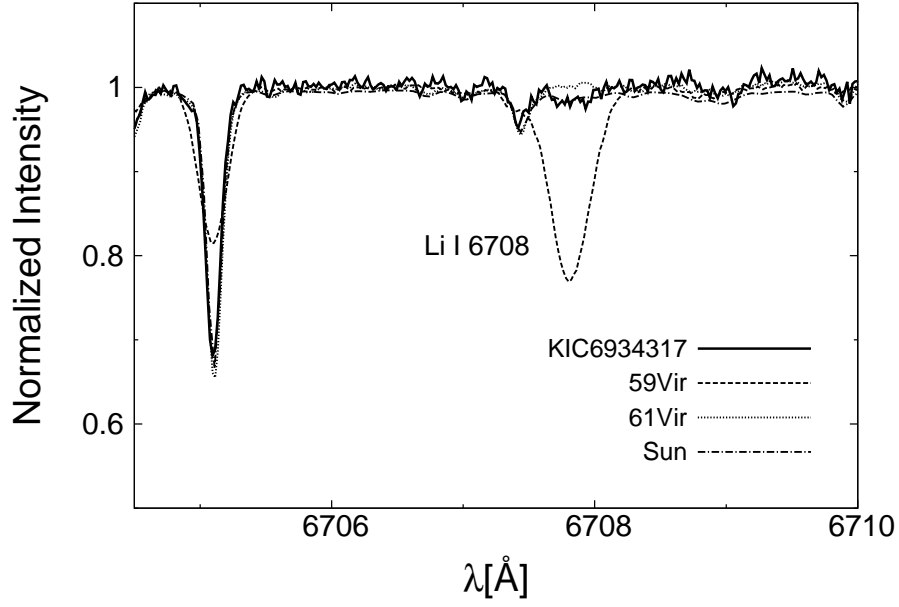}
  \end{center}
  \caption{\normalsize Normalized spectra of KIC6934317 (solid line), 59Vir (dashed line), 61Vir (dotted line), and the
  Sun (dash-dotted line) around Li I 6708. 
  The spectrum of the Sun is the same one which we used in Figure 3.
 According to these spectra, 59Vir has a deep line of Li, and the Li line of 61Vir is absent while that of the Sun
 and KIC6934317 are very weak and just visible against the noise.
 In Table 2 and 3, we show the values of lithium abundance (A(Li)) of KIC6934317, 59Vir, and 61Vir measured in this 
 study on the basis of the
  method described in \citet{KTakeda2005}.}
\label{fig5}
\end{figure}

\newpage

\begin{figure}
  \begin{center}
    \FigureFile(150mm,150mm){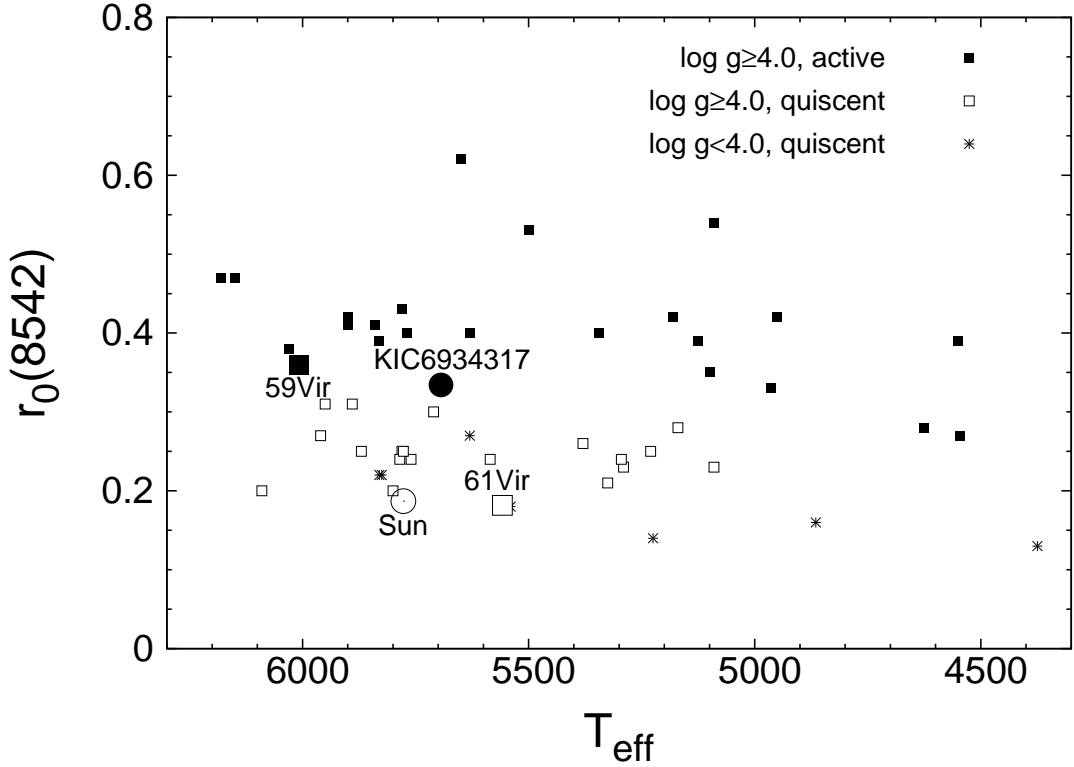}
  \end{center}
  \caption{\normalsize $r_{0}(8542)$ (Residual flux normalized by the continuum at the line core of Ca II 8542.) as a function of 
  effective temperature ($T_{\mathrm{eff}}$) of the stars. In this figure, we plot the results of this observation
  (large symbols), and 
  the data in \citet{Chmielewski2000}(small symbols). Small filled squares represent active dwarf stars (log \textit{g}$\geq$4.0), small open squares 
  indicate quiescent dwarfs ($\log g$$\geq$4.0), and small asterisks display evolved (giant or 
  sub-giant) stars which are all considered to be quiescent ($\log g$$<$4.0).
  There appears to be a clear dividing gap between active and quiescent stars which show some 
  slope with effective temperature (e.g., \cite{Foing1989}; \cite{Chmielewski2000}).
  The large filled square indicates 59Vir, the large open square represents 61Vir, the large open circle with a 
  point in the center of the circle is the Sun, and the large filled circle is KIC6934317.
  The position of KIC6934317, 59Vir, 61Vir, and the Sun in this figure indicate that 59Vir is an active star, and 61Vir and the Sun 
  are not active stars, and also suggest that KIC6934317 is a relatively active star.}\label{fig6}
\end{figure}

\newpage

\begin{figure}
  \begin{center}
    \FigureFile(150mm,150mm){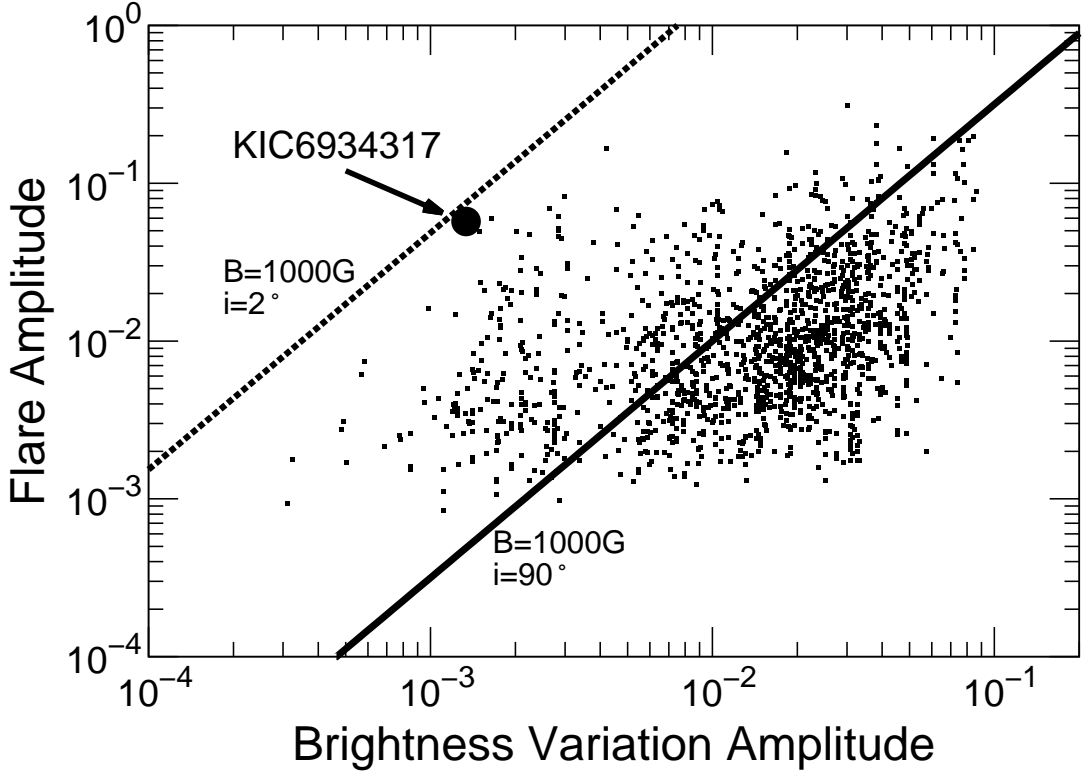}
  \end{center}
  \caption{\normalsize Scatter plot of superflare amplitude as a function of the amplitude of the brightness variation. The 
  data (small filled squares) are taken from \citet{YNotsu2013}. The solid line corresponds to the analytic relation between 
  the stellar brightness variation amplitude (corresponding to the starspot coverage) 
  and superflare amplitude (corresponding to the energy of superflare) obtained 
  from \citet{YNotsu2013} for $B$=1000G. The dashed line corresponds to the same relation in the case of 
  nearly pole-on (\textit{i}=2.0 deg) 
  for $B$=1000G, assuming that the brightness variation of a star become small as the inclination of the star become small. These lines 
  are considered to give an upper limit for the flare amplitude in each inclination. The large filled circle is KIC6934317 (The 
  brightness variation amplitude of KIC6934317 is about 0.1\%, and the amplitude 
  of the largest superflare which KIC6934317 exhibits in our data is 
  about 5.7\% of the brightness amplitude of the star.). This figure suggests that the inclination angle of KIC6934317 is 
  small (nearly pole-on) and that this star has large starspots. }\label{fig7}
\end{figure}


\end{document}